**Quantum mechanical and quasiclassical investigation of the time domain nonadiabatic dynamics of $NO_2$ close to the bottom of the $X^2A_1$-$A^2B_2$ conical intersection**

Michaël SANREY and Marc JOYEUX[a]

*Laboratoire de Spectrométrie Physique (CNRS UMR5588),*

*Université Joseph Fourier, BP 87, 38402 St Martin d'Hères, France*

**Abstract** : We use the effective Hamiltonian that we recently fitted against the first 306 experimentally observed vibronic transitions of $NO_2$ [J. Chem. Phys. 119, 5923 (2003)] to investigate the time domain nonadiabatic dynamics of this molecule on the coupled $X^2A_1$ and $A^2B_2$ electronic states, using both quantum mechanical and quasiclassical techniques. From the quantum mechanical point of view, we show that the transfer of population to the electronic ground state originating from a wave packet launched on the excited state occurs in a stepwise fashion. The evolution of wave packets launched on the electronic ground state is instead more complex because the crossing seam is located close to the bottom of the electronic excited state. We next use the mapping formalism, which replaces the discrete electronic degrees of freedom by continuous ones, to obtain a classical description of the coupled electronic states. We propagate gaussian swarms of trajectories to show that this approach can be used to calculate the populations in each electronic state. We finally propose a very simple trajectory surface hopping model, which assumes that trajectories have a constant probability to jump onto the other state in a particular region of the phase space and a null hopping probability outside from this region. Quasiclassical calculations show that this model enables a precise estimation of complex quantities, like for example the projection of the instantaneous probability density on given planes.

(a) email : Marc.JOYEUX@ujf-grenoble.fr

# I – Introduction

Vibronic interactions between different electronic potential energy surfaces are quite common in polyatomic molecules. If the electronic states are well separated in energy, they can be efficiently decoupled by applying the Born-Oppenheimer approximation and treating all the remaining small nonadiabatic effects within the framework of perturbation theory. As a result, the dynamics on the coupled surfaces is not very different from the dynamics on the isolated surfaces. In contrast, if the potential energy surfaces come close together, then the Born-Oppenheimer approximation breaks down and new phenomena, like radiationless decay, unimolecular reactions, electron transfers, *etc...* , become possible.

Triatomic molecules are the simplest systems, in which the most usual nonadiabatic effects, like the Renner-Teller effect, conical intersections and intersystem-crossing vibronic couplings, can take place. Amongst them, the $NO_2$ molecule has received particular attention, both from the experimental (see Refs. [1-4] and references therein) and theoretical (see Refs. [5-16] and references therein) points of view, so that it is now "the best known example where the effect of a conical intersection can be experimentally observed and calculated" [3]. Most of these studies were however performed in the energy domain because they were aimed at understanding the highly dense spectral lines in the optical spectra. The unimolecular time-dependent dynamics of $NO_2$ has been much less investigated [7,8,10,15]. In one of the first studies performed in the time domain [7], it was found that, after excitation in the upper electronic adiabatic surface, the system decays very rapidly (within less than one vibrational period) to the lower adiabatic state and then moves mainly on this surface. This led the authors to the conclusion that "the ground state nuclear dynamics can be understood without investigating higher electronic states" by using the adiabatic ground state surface. However, they also pointed out that this result is due to the fact that the coupling strength λ was assumed to be very strong ($\lambda$=2250 cm$^{-1}$) and that different regimes can be generated by varying $\lambda$. This remark is of fundamental importance, since it was shown more recently, on the basis of the comparison between experimental and calculated photoelectron spectra of $NO_2^-$ [9] and *ab initio* calculations [10], that the coupling strength is most certainly substantially smaller than the value used in Ref. [7]. Subsequent studies based on moderate values of the diabatic coupling ($\lambda \approx$700 cm$^{-1}$ in Ref. [10]) indeed displayed a less monotonous and richer behaviour of time-dependent quantities [10,15].

Still more recently, we derived an effective model for the $X^2A_1$-$A^2B_2$ conical intersection of $NO_2$ [16] by fitting its 30 parameters against the energies of 306 experimentally observed levels up to 11800 cm$^{-1}$ above the quantum mechanical ground state (there is only one missing level in the experimental spectrum). We obtained a coupling strength $\lambda \approx$330 cm$^{-1}$, which is again smaller than the coupling assumed in Refs. [10,15], although still much too large to be handled perturbatively. The purpose of the present paper is to investigate the *quantum mechanical* time domain dynamics of this model and to propose a *classical* model for the nonadiabatic dynamics of $NO_2$ close to the bottom of the $X^2A_1$-$A^2B_2$ conical intersection.

The remainder of this article is organized as follows. The effective model is briefly described in Sect. II for the sake of completeness. Quantum mechanical results, obtained from the "exact" propagation of gaussian wave packets are presented and discussed in Sect. III. These results are backed up in Sect. IV by the analysis of the dynamics of gaussian swarms of classical trajectories that (i) are governed by a classical scalar Hamiltonian obtained by mapping the discrete electronic degree of freedom on continuous variables [17-20], or (ii) evolve on decoupled diabatic electronic states and have a given probability to hop onto the other state in a certain region of the phase space.

## II – The effective model

The effective model used in this work is the one derived in Ref. [16], except that we henceforth neglect the vibrational resonance in the electronic ground state. Although experimentally ascertained, this resonance is so weak that the average error in computed energies increases only by a small fraction of a cm$^{-1}$ when it is disregarded. The diabatic Hamiltonian of the coupled system is thus of the form

$$\mathbf{H} = \begin{pmatrix} H_e & H_c \\ H_c & H_g \end{pmatrix}, \tag{2.1}$$

where $H_g$ and $H_e$ are the scalar Hamiltonians for the diabatic ground and excited states and $H_c$ is the diabatic coupling. In the following we use, in agreement with standard spectroscopic notations, indexes 1 to 3 to label quantities related, respectively, to symmetric stretch, bend and antisymmetric stretch. We furthermore denote by $(p_k, q_k)$ and $(p'_k, q'_k)$ the sets of dimensionless normal coordinates in the ground and excited states, and by $n_k = (p_k^2 + q_k^2)/2$ and $n'_k = (p'^2_k + q'^2_k)/2$ the corresponding action integrals. Since we neglect the weak vibrational resonance in the ground state, $H_g$ and $H_e$ are just polynomial functions of the action integrals

$$
\begin{aligned}
H_g &= \sum_{i=1}^{3} \omega_i n_i + \sum_{i \le j} x_{ij} n_i n_j + \sum_{i \le j \le k} y_{ijk} n_i n_j n_k + \sum_{i \le j \le k \le m} z_{ijkm} n_i n_j n_k n_m \\
H_e &= E'_0 + \sum_{i=1}^{3} \omega'_i n'_i + \sum_{i \le j} x'_{ij} n'_i n'_j
\end{aligned} \tag{2.2}
$$

(see Table I of Ref. [16] for numerical values of the parameters). The order of the Dunham expansion is smaller for $H_e$ than for $H_g$ because experimental data for the electronic excited state is much sparser than for the ground state. The diabatic coupling $H_c$ is taken to be the only first order term authorized by symmetry, that is

$$H_c = \lambda q_3 \ .$$ (2.3)

As we showed using canonical perturbation theory [21,22], it is not easy to extract information on $H_c$ from spectroscopic data because the effect of $H_c$ on line positions is to a large extent similar to that of anharmonicities inside each surface. Therefore, the only information on $H_c$ that could be derived from the experimental spectrum is the value $\lambda \approx 332 \pm 19$ cm$^{-1}$. At last, position coordinates $\mathbf{q} = (q_1, q_2, q_3)$ and $\mathbf{q'} = (q_1', q_2', q_3')$ are related through

$$\mathbf{q}' = \mathbf{A}\,\mathbf{q} + \mathbf{B} = \begin{pmatrix} 0.899 & -0.532 & 0 \\ 0.301 & 0.906 & 0 \\ 0 & 0 & 0.693 \end{pmatrix} \begin{pmatrix} q_1 \\ q_2 \\ q_3 \end{pmatrix} + \begin{pmatrix} 1.100 \\ -5.730 \\ 0 \end{pmatrix} ,$$ (2.4)

while momenta $\mathbf{p} = (p_1, p_2, p_3)$ and $\mathbf{p'} = (p_1', p_2', p_3')$ consequently satisfy

$$\mathbf{p}' = {}^t\mathbf{A}^{-1}\,\mathbf{p} \ .$$ (2.5)

Eq. (2.4) was obtained, by using standard techniques for the calculation of normal mode coordinates, from (i) the firmly established equilibrium geometry in the electronic ground state, (ii) the more indirectly determined geometry in the electronic excited state [4], (iii) experimentally determined harmonic frequencies, and (iv) ab initio off-diagonal quadratic force constants for X$^2$A$_1$ and A$^2$B$_2$ [23]. Note that the minimum of the electronic excited state, at about 10200 cm$^{-1}$ above that of the ground state, is located very close to the crossing seam. This will play an important role in the propagation of wave packets.

Since $\lambda$ is of the order of several hundreds of cm$^{-1}$, the diabatic coupling between harmonic vectors separated by several tens of thousands of cm$^{-1}$ still has a sizable effect on the energy of each eigenstate. We therefore proposed in Ref. [16] to use optimal bases obtained from perturbation theory in order to reduce strongly the size of the Hamiltonian matrix and to speed up quantum mechanical calculations. This method was used to generate the set of converged eigenstates on which each wave packet was initially projected (see Sect. III).

## III – Quantum mechanical description

In this work, we did not try to mimic a precise experiment, in the sense that the initial wave packet was not chosen as the Franck-Condon excitation of the vibronic ground state. In order to get a more general understanding of the time domain dynamics on the coupled surfaces, we instead placed wave packets at several positions in the ground and excited states and let them evolve with time. These wave packets were assumed to be initially of minimum uncertainty with respect to the normal coordinates of the diabatic state on which they were launched. More explicitly, wave packet launched on the electronic excited state were of the form

$$\Phi_{\overline{\mathbf{p}}',\overline{\mathbf{q}}'}(\mathbf{q}',t=0) = \begin{pmatrix} \prod_{k=1}^{3} \pi^{-1/4} \exp\left\{ i\,\overline{p}'_k q'_k - \frac{1}{2}\left(q'_k - \overline{q}'_k\right)^2 \right\} \\ 0 \end{pmatrix}, \tag{3.1}$$

where $\left(\overline{\mathbf{p}}',\overline{\mathbf{q}}'\right)$ denotes the coordinates and momenta of the center of the wave packet at $t=0$, while wave packets launched on the electronic ground state were of the form

$$\Phi_{\overline{\mathbf{p}},\overline{\mathbf{q}}}(\mathbf{q},t=0) = \begin{pmatrix} 0 \\ \prod_{k=1}^{3} \pi^{-1/4} \exp\left\{ i\,\overline{p}_k q_k - \frac{1}{2}\left(q_k - \overline{q}_k\right)^2 \right\} \end{pmatrix}, \tag{3.2}$$

with similar notations. Note that we use matrix notations throughout this paper, so that the zeroes that appear in the second row of Eq. (3.1) and in the first row of Eq. (3.2) simply mean that the corresponding wave packets have no component at time $t=0$ on the electronic ground and excited states, respectively. These wave packets were projected on the eigenvectors

$$\mathbf{\Psi}_j = \begin{pmatrix} \Psi_j^e \\ \Psi_j^g \end{pmatrix} \qquad (3.3)$$

of the coupled system, in order to expand them according to

$$\mathbf{\Phi}(t=0) = \sum_j c_j \mathbf{\Psi}_j \qquad (3.4)$$

(for the sake of simpler notations, explicit reference to momenta and position coordinates is dropped). Wave packets at time $t$ then satisfy

$$\mathbf{\Phi}(t) = \begin{pmatrix} \Phi^e(t) \\ \Phi^g(t) \end{pmatrix} = \sum_j c_j \exp(-iE_j t) \mathbf{\Psi}_j = \begin{pmatrix} \sum_j c_j \exp(-iE_j t) \Psi_j^e \\ \sum_j c_j \exp(-iE_j t) \Psi_j^g \end{pmatrix}, \qquad (3.5)$$

where $E_j = \langle \mathbf{\Psi}_j | \mathbf{H} | \mathbf{\Psi}_j \rangle$ denotes the energy of vibronic state $j$. The expressions for most instantaneous quantities follow straightforwardly from Eq. (3.5). For example, the populations $P^\xi(t)$ in each electronic state $\xi$ ($\xi = g, e$) obey

$$P^\xi(t) = \langle \Phi^\xi(t) | \Phi^\xi(t) \rangle \qquad (3.6)$$

(note that $P^g(t) + P^e(t) = 1$), while the average of any operator $A$ in a given electronic state $\xi$ is obtained according to

$$\overline{A}^\xi(t) = \frac{1}{P^\xi(t)} \langle \Phi^\xi(t) | A | \Phi^\xi(t) \rangle \qquad (3.7)$$

(note that $\overline{p}_k^g(0) = \overline{p}_k$, $\overline{q}_k^g(0) = \overline{q}_k$, $\overline{p}_k'^e(0) = \overline{p}_k'$, $\overline{q}_k'^e(0) = \overline{q}_k'$, etc...). For the purpose of visualizing the evolution of the wave packet, we also plotted the projection of the instantaneous probability density of each component of the wave packet on the $(q_1, q_2)$ plane, that is,

$$\rho^{\xi}(q_1, q_2, t) = \int \left| \Phi^{\xi}(t) \right|^2 dq_3 \quad . \tag{3.8}$$

The remainder of this section is devoted to a discussion of the results obtained by the procedure sketched above. Illustrations will focus on the evolution of two wave packets, which are representative of all the results we obtained. The first one is initially placed with no impulsion ($\overline{\mathbf{p}}' = \overline{\mathbf{p}} = (0,0,0)$) on the electronic excited state at position $\overline{\mathbf{q}} = (1.0, 3.0, 1.5)$ (use Eq. (2.4) to obtain the corresponding value of $\overline{\mathbf{q}}'$), while the second one is initially placed with no impulsion on the electronic ground state at position $\overline{\mathbf{q}} = (1.25, 5.5, 1.0)$. In both cases, the energy at $t = 0$ of the center of the wave packet is comprised between 13000 and 13500 cm$^{-1}$ above the minimum of the electronic ground state.

The first result, illustrated in Fig. 1, is that the component of the wave packet that remains on the initial electronic state essentially follows, at least for the first tens of femtoseconds, the trajectory of the initial center of the wave packet on the decoupled electronic state. This is clearly seen in Fig. 1, which shows the time evolution of the coordinates $q_k$ for the trajectory of $H_g$ with initial coordinates $\mathbf{p} = (0,0,0)$ and $\mathbf{q} = (1.25, 5.5, 1.0)$ (dash-dotted lines) and the average values $\overline{q}_k^g(t)$ for the wave packet launched on $H_g$ (solid lines). The classical and quantum mechanical curves are very close at short times. The dephasing between the two sets of curves however increases almost linearly with time. This is due to the fact that the diabatic coupling induces anharmonicities in both electronic states [21,22]. Therefore, the frequencies experienced by the quantum mechanical wave packet, which evolves on the coupled electronic states, are slightly but noticeably different from the frequencies experienced by the classical trajectory, which evolves on the decoupled ground state. Moreover, the spreading of the quantum wave packet and/or the energy and probability that are transferred to the other electronic surface damp the amplitude of the oscillations of the $\overline{q}_k^g(t)$ curves.

The essentially classical behavior of the component of the wave packet that remains on the initial electronic state is also clearly seen in Figs. 2 and 3. These figures show the projection of the probability densities on the $(q_1, q_2)$ plane, $\rho^\xi(q_1, q_2, t)$, for the wave packet launched on the electronic excited state (Fig. 2) and the one launched on the electronic ground state (Fig. 3). The top plots show the densities for the component of the wave packet on the electronic excited state, that is $\rho^e(q_1, q_2, t)$, while the bottom plots show the densities for the component on the electronic ground state , that is $\rho^g(q_1, q_2, t)$. The vignettes correspond roughly to the times where the wave packet is at maximum bending deviation in the initial surface, but the Lissajou curves that the wave packet approximately follows in the initial surface are easy to imagine.

The transitions from one electronic state to the other one, as well as the behavior of the wave packet on the initially unpopulated state, are less straightforward. A clear picture seems however to emerge when comparing the "trajectories" in Fig. 1 to the density probabilities in Figs. 2 and 3 and the population plots in Fig. 4. In this later figure, the solid lines represent the time evolution of the population $P^\xi(t)$ in the initial electronic state for the wave packet launched on the electronic excited state (top plot) and the one launched on the ground state (bottom plot). It is seen that the transfer of density from the initially populated electronic state to the other one occurs in a stepwise fashion. The plateaux are flat for the wave packet launched on the electronic excited state, while they have an almost regular positive slope for the wave packet launched on the electronic ground state. Comparison of the various figures indicates that the sharp transfers of density take place when the initial wave packet is located in a region of the configuration space close to the crossing seam of the two diabatic states. As long as the initial wave packet moves in this "intersection region", which will be characterized more precisely in Sect. IV, it transfers probability and energy to the other surface. After some femtoseconds, the initial wave packet however leaves the intersection region and there are henceforth two almost independent components, one on each electronic state.

As is seen in Fig. 2 (vignettes at $t = 25$ fs), the two components emanating from the wave packet launched on the electronic excited surface then oscillate independently in their respective wells. No density is transferred during this round trip, which corresponds to the first plateau in the top plot of Fig. 4. When the excited state component comes back in the intersection region, it transfers a second contribution to the ground state. The associated population decay in the excited state is reflected by the second sharp decrease in the top plot of Fig. 4. Of course, the component on the electronic ground state simultaneously transfers some probability to the excited state, but the relative populations at that time are such that this back-transfer is almost unnoticeable in the plots. The vignettes at $t = 45$ fs (Fig. 2) show that the two contributions that have successively been transferred to the electronic ground state interfere, leading to clear fringes in the density plot. The components on each electronic surface then oscillate again (vignettes at $t = 70$ fs) and the whole process is repeated till the transfers in both directions equilibrate at around 180 fs (see the insert in the top plot of Fig. 4). Quite satisfactorily, the results illustrated in Fig. 2 and in the top plot of Fig. 4 are in qualitative agreement with corresponding ab initio results (see Figs. 5 and 6 of Ref. [10]).

Things are more complex for the wave packet launched on the electronic ground state. The reason is that the seam of the two electronic states, as well as the intersection region, are located very close to the bottom of the excited state. Therefore, the component on the excited state oscillates with very reduced amplitude and remains close to or in the intersection region. This has two consequences. First, the contributions transferred to the excited state interfere with themselves (see the vignette at $t = 20$ fs in Fig. 3), in addition to interfering with contributions transferred at different times (see the vignettes at $t = 48$ and $t = 70$ fs in Fig. 3).
Moreover, since the excited state component remains localized in the intersection region, it
continuously transfers probability back to the electronic ground state. This is the reason why
the population in the electronic ground state slightly but regularly increases during the
plateaux : the ground state component oscillates outside from the intersection region (see the
vignettes at $t = 48$ and $t = 70$ fs) and is thus not able to transfer probability to the excited
state, while the excited state component remains localized in the intersection region and
continuously transfers probability back to the ground state.

This later conclusion was confirmed by calculations where wave packets were
propagated using first order time dependent perturbation theory. This indeed led to very
regular steps and flat plateaux in the time evolution of electronic populations, whatever the
state on which the wave packet was initially launched. Since first order perturbation theory
takes into account the transfer from the initial state to the other one, but not the transfer back
to the initial state, this confirms that the positive slopes of the plateaux in the bottom plot of
Fig. 4 are indeed due to the back transfers.

The description of the time domain dynamics of $NO_2$ close to the bottom of the $X^2A_1$-
$A^2B_2$ conical intersection, which is formulated above, will now be backed by the analysis of
the dynamics of gaussian swarms of classical trajectories.

**IV - Quasiclassical description**

The correspondence of quantum and classical dynamics is an active field of research in
molecular physics [24-30]. With this respect, the analysis of the dynamics on several coupled
electronic states however represents a more complex problem than, for example, the
description of abrupt changes in vibrational wave functions in terms of bifurcations of the classical periodic orbits [27,29,30], because the electronic degrees of freedom have no obvious classical analogue. Two approaches have traditionally been used to incorporate quantum degrees of freedom into a classical formulation. The first one, which was pioneered by Landau [31], Zener [32] and Stückelberg [33], incorporates nonadiabatic transitions between coupled electronic states through a surface-hopping procedure [34-36]. The second approach is instead based on the replacement of the discrete degrees of freedom by continuous ones, either at the classical [37-42] or quantum mechanical [17-20,43,44] levels. The description obtained from the quantum mechanical transformation of discrete onto continuous degrees of freedom has been called the "mapping" approach by its authors.

We present below the results obtained with (i) the mapping approach, and (ii) a surface-hopping approach in the diabatic representation derived from the mapping Hamiltonian. We start with the mapping approach, which is simpler to implement.

According to Refs. [17-20], the mapping Hamiltonian associated with the quantum mechanical Hamiltonian of Eq. (2.1) can be written in the form

$$H = \frac{1}{2}\left(P_g^2 + Q_g^2 - 1\right)H_g + \frac{1}{2}\left(P_e^2 + Q_e^2 - 1\right)H_e + \left(P_g P_e + Q_g Q_e\right)H_c \,, \qquad (4.1)$$

where the $\left(P_\xi, Q_\xi\right)$ are sets of conjugate "electronic" coordinates that describe the electronic ground $(\xi = g)$ and excited $(\xi = e)$ states. In order to get the quasiclassical equivalents of the electronic state populations $P^\xi(t)$ plotted in Fig. 4, one just needs to propagate classically swarms of trajectories, which at time $t = 0$ are normally distributed (with $1/\sqrt{2}$ variance) around the center of the quantum wave packet, and take the average of $\frac{1}{2}(P_\xi^2 + Q_\xi^2 - 1)$ over all the trajectories. The results obtained with swarms of $2 \ 10^4$ trajectories are shown as dot-dot-dot-dashed lines in Fig. 4. It is seen that the classical mapping Hamiltonian reproduces correctly the stepwise decrease of the electronic populations in the interval 0-180 fs. At longer times, it slightly underestimates the excited state population for the wave packet launched on the excited state, but results are still qualitatively correct.

However, the essential drawback of the mapping approach is that the vibronic trajectories evolve on states that are linear combinations of the electronic ground and excited states but rarely coincide with them. Stated in other words, $\frac{1}{2}(P_\xi^2 + Q_\xi^2 - 1)$ may take any value comprised between $-\frac{1}{2}$ and $\frac{3}{2}$ instead of just 0 and 1. Moreover, simple approximations, like for example considering that a given trajectory is on the electronic ground state if $\frac{1}{2}(P_e^2 + Q_e^2 - 1) < \frac{1}{2}$ and on the excited state if $\frac{1}{2}(P_e^2 + Q_e^2 - 1) > \frac{1}{2}$, just lead to completely meaningless results. Therefore, the standard mapping approach is not adapted to the quasiclassical calculation of many state-specific quantities, like for example the probability densities shown in Figs. 2 and 3. To this end, we could have implemented the "decay of mixing" procedure in the mapping formalism (see Ref. [45] and references therein). We could also have adiabatized the quantum mechanical Hamiltonian of Eq. (2.1) to use, for example, Tully's fewest switches trajectory surface-hopping procedure [34-36]. We instead preferred to develop a somewhat different classical model, which we hope is particularly easy to grab. According to this model, which is based on the quantum mechanical observations in Sect. III, a given trajectory may hop onto the other electronic state only when it travels inside a certain region of the phase space called the "intersection region". The probability to jump onto the other state is constant inside the intersection region and null outside. The notion of "intersection region" could be given a theoretical foundation by analyzing in some more detail the mapping Hamiltonian of Eq. (4.1). Indeed, by introducing the action-angle variables $\left(I_\xi, \phi_\xi\right)$ associated with the "electronic" degrees of freedom

$$Q_\xi = \sqrt{2I_\xi} \cos \phi_\xi$$
$$P_\xi = -\sqrt{2I_\xi} \sin \phi_\xi$$

(4.2)

(note that $\frac{1}{2}(P_\xi^2 + Q_\xi^2 - 1) = I_\xi - \frac{1}{2}$), one easily sees that

$$\frac{dI_g}{dt} = -\frac{dI_e}{dt} = 2H_c \sqrt{I_g I_e} \sin\left(\phi_g - \phi_e\right) \qquad (4.3)$$

and

$$\frac{d}{dt}\left(\phi_g - \phi_e\right) = H_g - H_e + H_c \left(\sqrt{\frac{I_e}{I_g}} - \sqrt{\frac{I_g}{I_e}}\right)\cos\left(\phi_g - \phi_e\right). \qquad (4.4)$$

Most of the time, $\left|H_g - H_e\right|$ is very large, so that the electronic phase $\phi_g - \phi_e$ oscillates very rapidly and the fluctuations of $I_g$ and $I_e$ average to zero. The only exception occurs when the electronic phase is almost stationary, that is, when the right-hand side of Eq. (4.4) is close to zero. For trajectories located on one of the electronic states, $\left(I_g, I_e\right)$ is equal to either $\left(\frac{1}{2}, \frac{3}{2}\right)$ or $\left(\frac{3}{2}, \frac{1}{2}\right)$ and the stationary phase condition can be satisfied only in the region of the phase space defined by

$$\left|H_g - H_e\right| \le \left(\sqrt{3} - \frac{1}{\sqrt{3}}\right)\left|H_c\right| . \qquad (4.5).$$

We could have considered the inequality of Eq. (4.5) as the definition of the intersection region. However, to take into account the fact that the electronic phase needs not be exactly stationary but must instead just evolve slowly, we multiplied the numerical prefactor in the right-hand side of Eq. (4.5) by approximately two and defined the intersection region as the region where the inequality

$$\left|H_g - H_e\right| \le 2\left|H_c\right| \qquad (4.6)$$

holds. At last, the hopping rates need only be adjusted in order that the time dependent populations obtained with the hopping approach match those obtained with the mapping or quantum mechanical ones. Note that these two rates govern essentially the average slope at short times as well as the equilibrium value at long times, *but not the stepwise decay*. Stated in other words, the succession of steps is due to the motion of the wave packets inside and outside from the intersection region and not to particular values of the hopping rates. Since quantum mechanical populations were available and our principal goal was to validate our description of the nonadiabatic dynamics in terms of the intersection region, we varied here the two rates till the time dependent populations obtained with the hopping model were sufficiently close to the quantum mechanical ones in the 0-180 fs time interval. For the swarm launched on the electronic excited state, we thus used $g \to e$ and $e \to g$ hopping rates of 0.38 and 0.20 fs$^{-1}$, respectively, while for the swarm launched on the electronic ground state, we used rates of 0.33 and 0.20 fs$^{-1}$, respectively. Of course, these quantities depend on the exact definition of the intersection region. For example, if it were defined according to Eq. (4.5) instead of Eq. (4.6), then one should use rates larger by (approximately) a factor 2.

The populations obtained with swarms of 2 10$^4$ trajectories are shown as dashed lines in Fig. 4. It is seen that they agree very well with quantum mechanical results in the interval 0-180 fs, which indicates that the definition of the intersection region in Eq. (4.6) is judicious. The result is particularly impressive for the swarm launched on the electronic ground state, since we initially had doubts concerning the ability of a classical procedure to mimic correctly the unceasing forth and back transfers between the two states. While still qualitatively correct, the agreement between quantum mechanical and hopping results is slightly less good at longer times ($t$>180 fs), but this is essentially due to the fact that we considered only the 0-180 fs interval for the adjustment of the hopping rates.

The projections of the probability densities on the $(q_1, q_2)$ plane, $\rho^\xi(q_1, q_2, t)$, obtained with swarms of 10$^6$ trajectories launched on $H_e$ and $H_g$ are shown in Figs. 5 and 6, respectively. These figures are the quasiclassical equivalents of Figs. 2 and 3, respectively. Except, of course, for interference effects, we again note a very good agreement between quantum mechanical and quasiclassical results, even for the swarm launched on the electronic ground state, for which the density transferred to the excited state oscillates in the intersection region.

**V - Conclusion**

We have presented a time-dependent dynamical study on the coupled $X^2A_1$ and $A^2B_2$ electronic states of $NO_2$, which is based on the effective Hamiltonian that we have recently fitted against the first 306 experimentally observed vibronic transitions of this molecule. We have shown that, for wave packets launched on the electronic excited state, the transfer of population to the ground state occurs in a stepwise fashion. The evolution of wave packets launched on the electronic ground state is instead more complex because the crossing seam is located close to the bottom of the electronic excited state. We have used the mapping formalism to obtain a classical description of the coupled electronic states and propagated gaussian swarms of trajectories to show that this approach can be used to calculate with good accuracy the populations in each electronic state. We have finally proposed a very simple trajectory surface hopping model, which nevertheless reproduces with very good accuracy more complex quantities, like for example the projection of the instantaneous probability density on given planes.

**FIGURE CAPTIONS**

**Figure 1** (color online): Dash-dotted lines : time evolution of the coordinates $q_k$ for the trajectory of $H_g$ with initial coordinates $\overline{\mathbf{p}} = (0,0,0)$ and $\overline{\mathbf{q}} = (1.25, 5.5, 1.0)$. Solid lines : time evolution of the average values $\overline{q}_k^{\,g}(t)$ for the wave packet launched on $H_g$ and initially centered at $\overline{\mathbf{p}} = (0,0,0)$ and $\overline{\mathbf{q}} = (1.25, 5.5, 1.0)$. Indexes 1 to 3 refer to the symmetric strech, the bend, and the antisymmetric stretch degrees of freedom, respectively.

**Figure 2** (color online) : Probability densities $\rho^{\xi}(q_1, q_2, t)$ for the wave packet launched on $H_e$ and initially centered at $\overline{\mathbf{p}} = (0,0,0)$ and $\overline{\mathbf{q}} = (1.0, 3.0, 1.5)$. Top plots show the probability $\rho^e(q_1, q_2, t)$ for the component of the wave packet on the electronic excited state, while bottom plots show the probability $\rho^g(q_1, q_2, t)$ for the component on the electronic ground state. Note that plot intensities are not normalized. The four pairs of vignettes correspond roughly to the times where the wave packet is at maximum bending deviation in the initial electronic surface. The thin lines are contours plots of the pseudo-potential energies at $q_3 = 0$. Contours range from 2000 to 18000 cm$^{-1}$ for $H_g$ and from 12000 to 18000 cm$^{-1}$ for $H_e$, with 2000 cm$^{-1}$ increments. The thicker line indicates the crossing seam of the two surfaces.

**Figure 3** (color online) : Probability densities $\rho^{\xi}(q_1, q_2, t)$ for the wave packet launched on $H_g$ and initially centered at $\overline{\mathbf{p}} = (0,0,0)$ and $\overline{\mathbf{q}} = (1.25, 5.5, 1.0)$. Top plots show the probability $\rho^e(q_1, q_2, t)$ for the component of the wave packet on the electronic excited state, while bottom plots show the probability $\rho^g\left(q_1,q_2,t\right)$ for the component on the electronic ground state. Note that plot intensities are not normalized. The four pairs of vignettes correspond roughly to the times where the wave packet is at maximum bending deviation in the initial electronic surface. The thin lines are contours plots of the pseudo-potential energies at $q_3 = 0$. Contours range from 2000 to 18000 cm$^{-1}$ for $H_g$ and from 12000 to 18000 cm$^{-1}$ for $H_e$, with 2000 cm$^{-1}$ increments. The thicker line indicates the crossing seam of the two surfaces.

**Figure 4** (color online) : Time evolution of the populations in the initial electronic state for quantum wave packets and swarms of 20000 classical trajectories launched (i) on $H_e$ and initially centered at $\overline{\mathbf{p}} = \left(0,0,0\right)$ and $\overline{\mathbf{q}} = \left(1.0,\,3.0,\,1.5\right)$ (top plot), and (ii) on $H_g$ and initially centered at $\overline{\mathbf{p}} = \left(0,0,0\right)$ and $\overline{\mathbf{q}} = \left(1.25,\,5.5,\,1.0\right)$ (bottom plot). Solid, dashed and dot-dot-dot-dashed lines show the results obtained from "exact" quantum mechanical propagation of the wave packet, the classical hopping model and the classical mapping model, respectively. The main plots focus on the short time dynamics (up to 180 fs), while the inserts show the populations up to 500 fs.

**Figure 5** (color online) : Same as Fig. 2, but obtained by propagating swarms of $10^6$ trajectories according to the trajectory surface hopping procedure described in Sect. IV.

**Figure 6** (color online) : Same as Fig. 3, but obtained by propagating swarms of $10^6$ trajectories according to the trajectory surface hopping procedure described in Sect. IV.

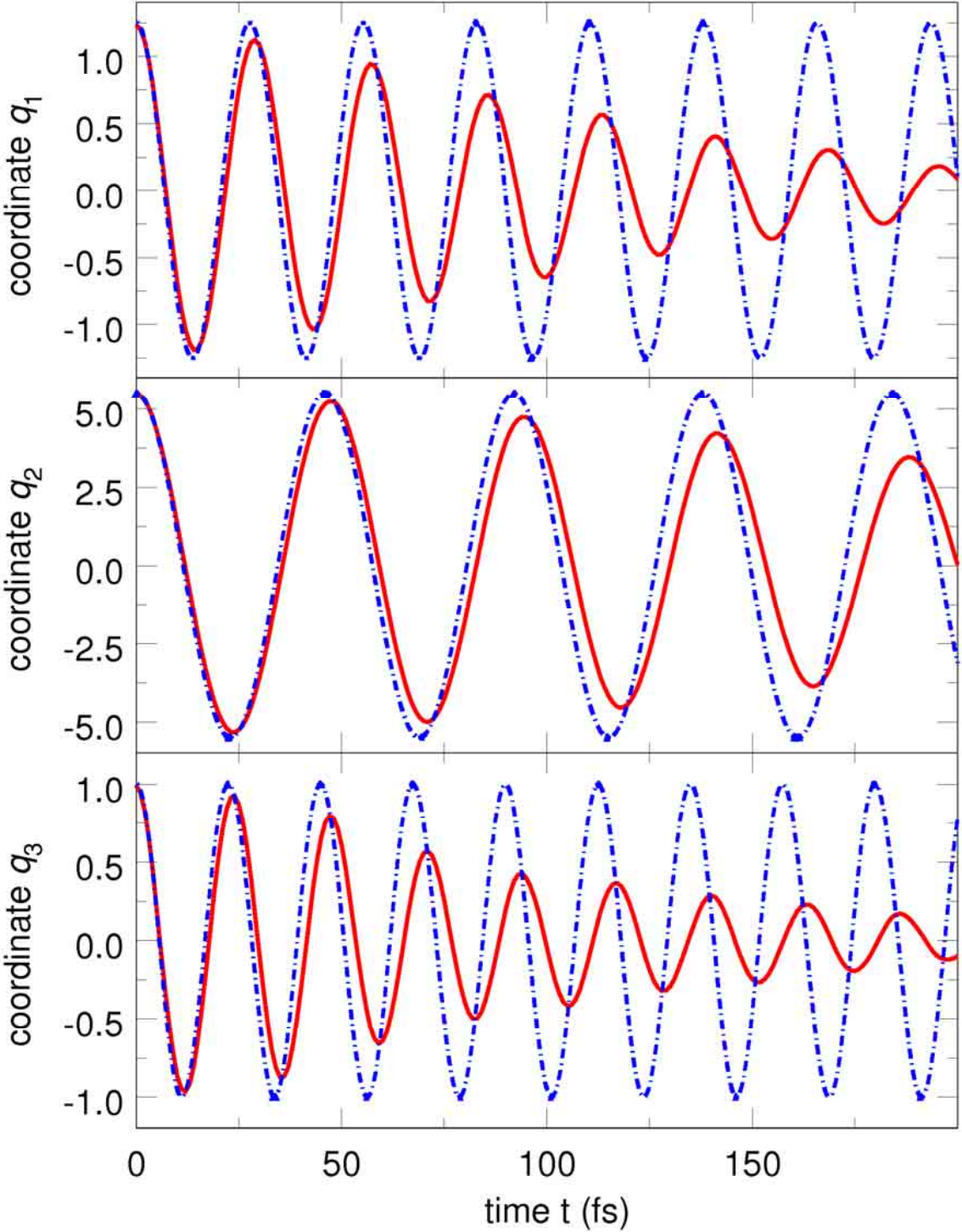

**FIGURE 2**

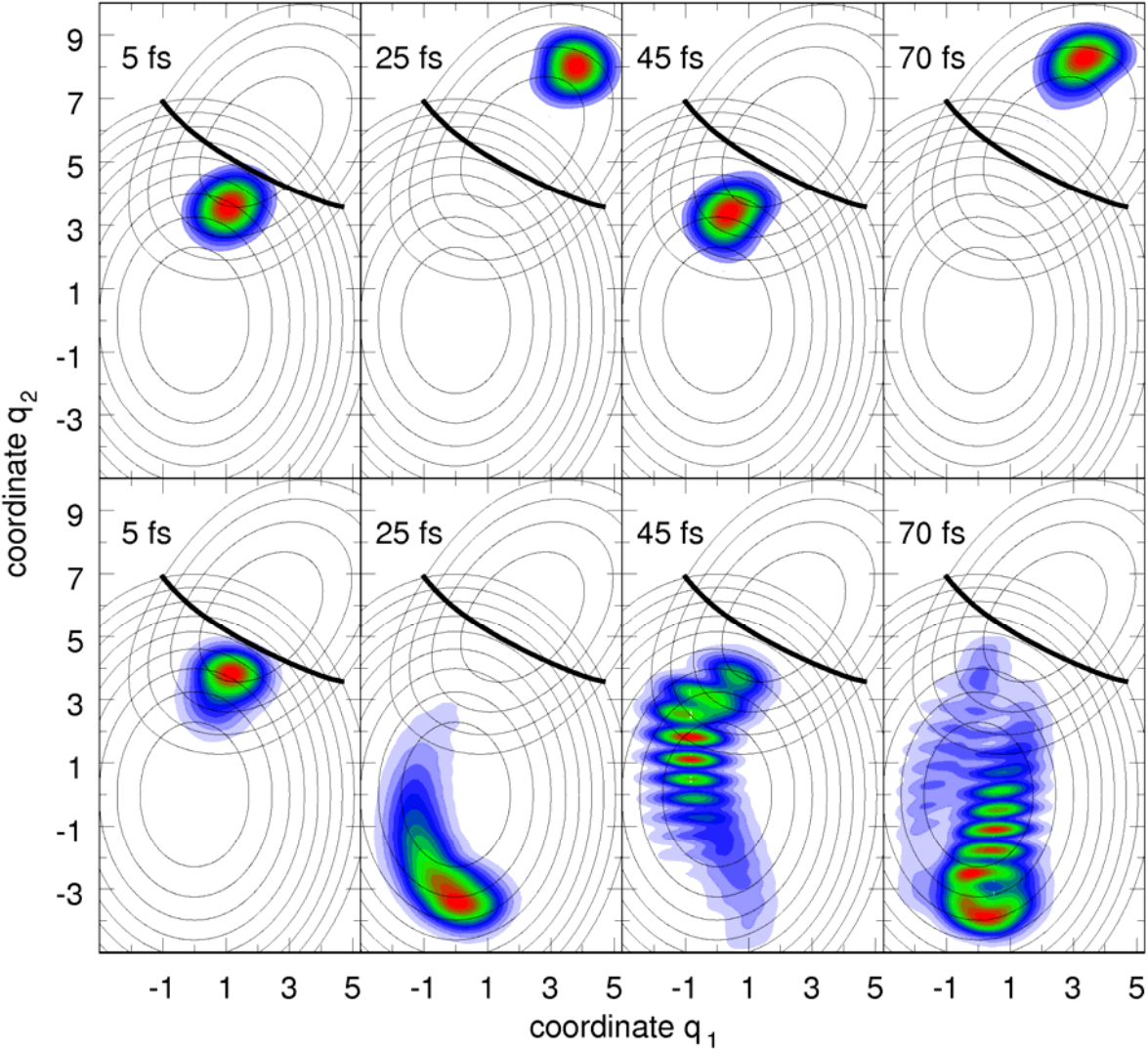

**FIGURE 3**

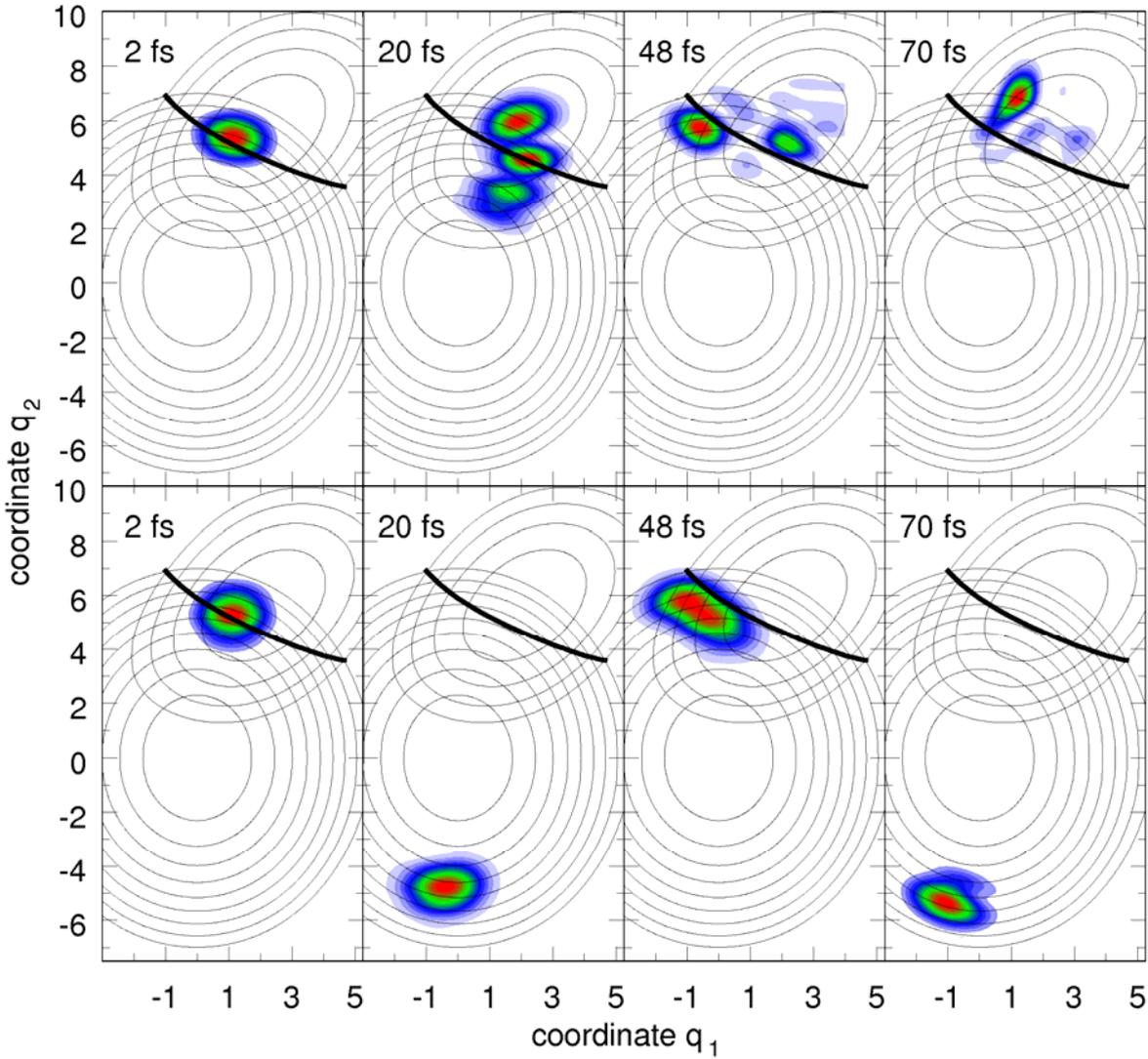

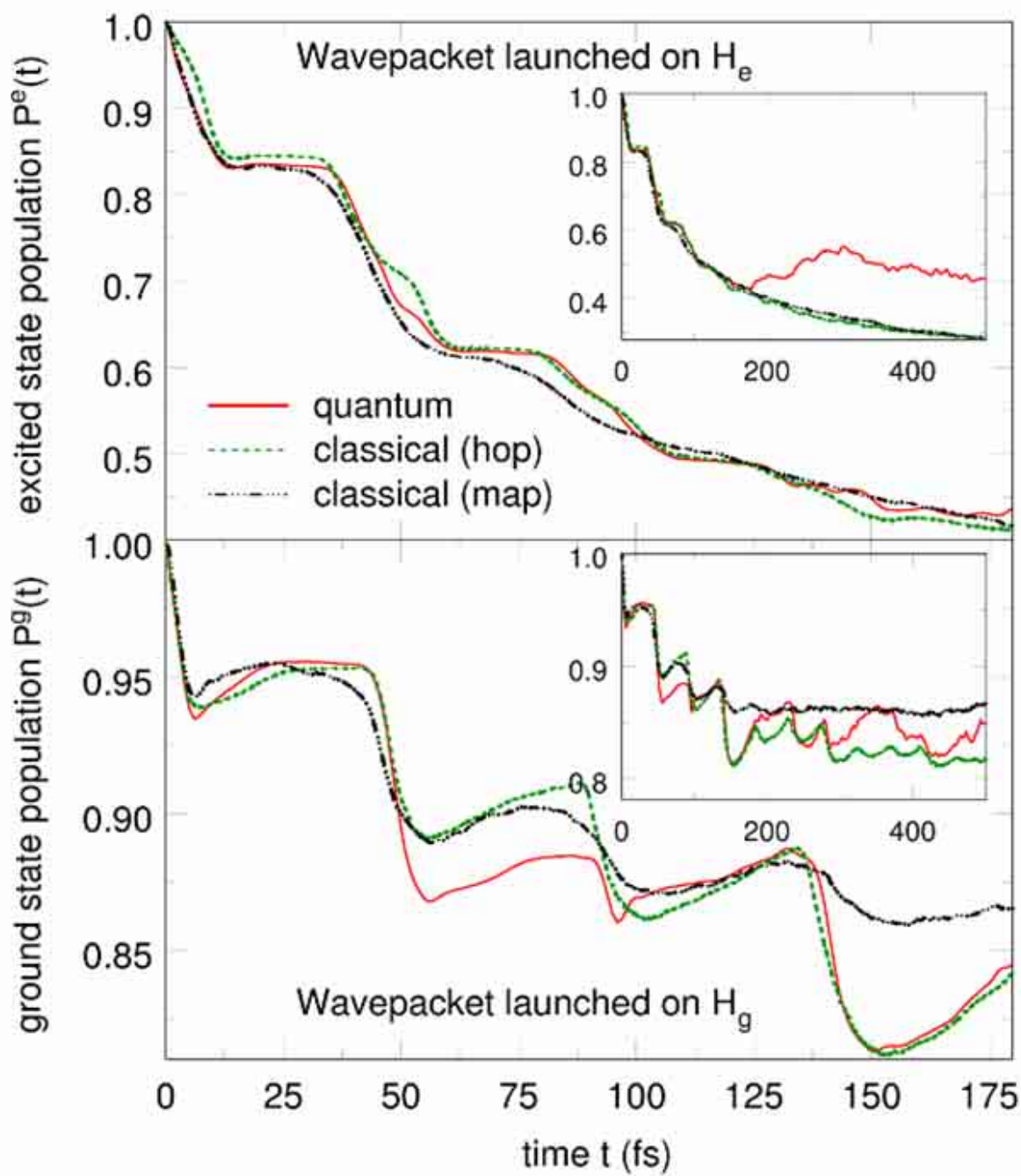

**FIGURE 5**

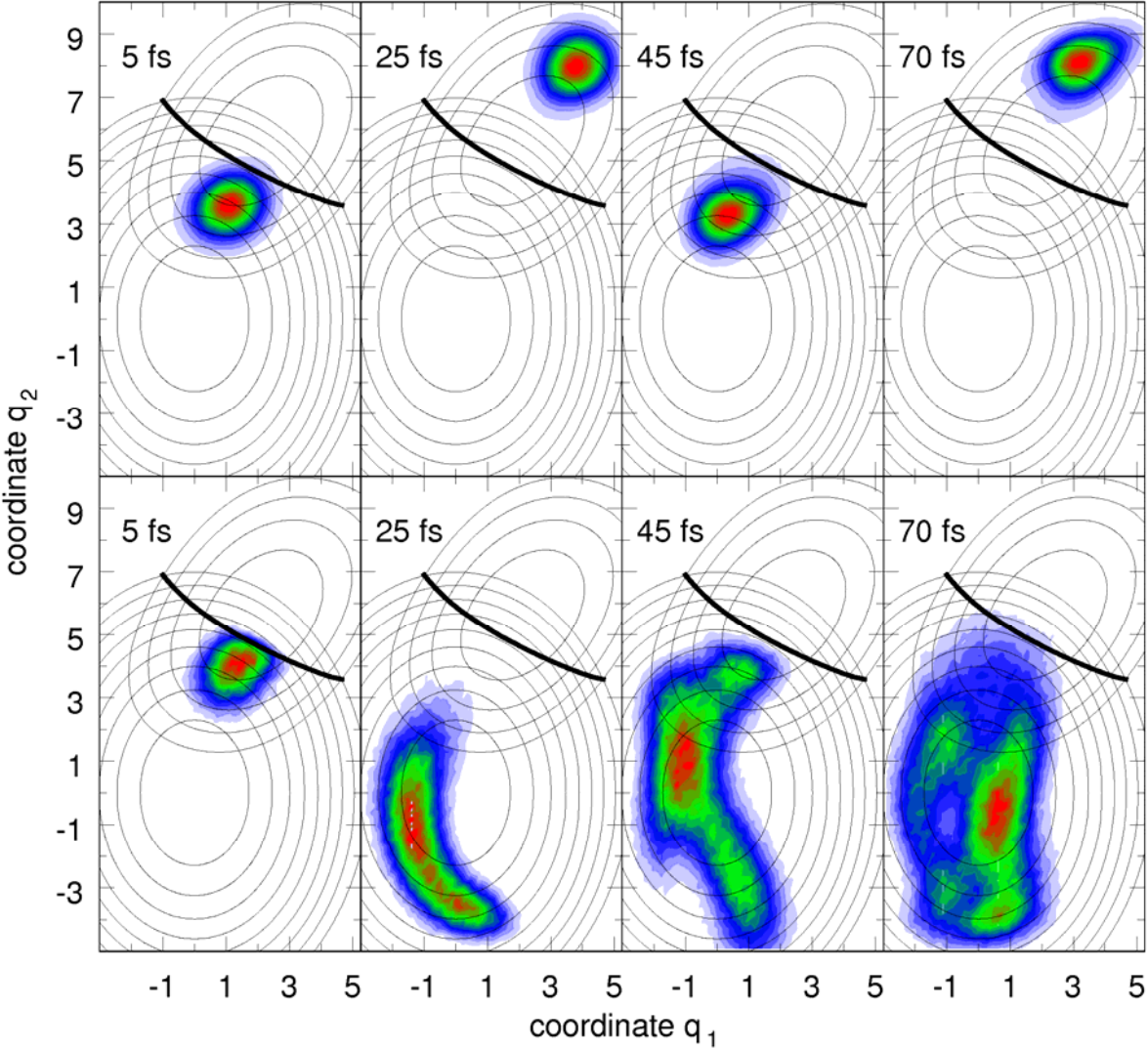

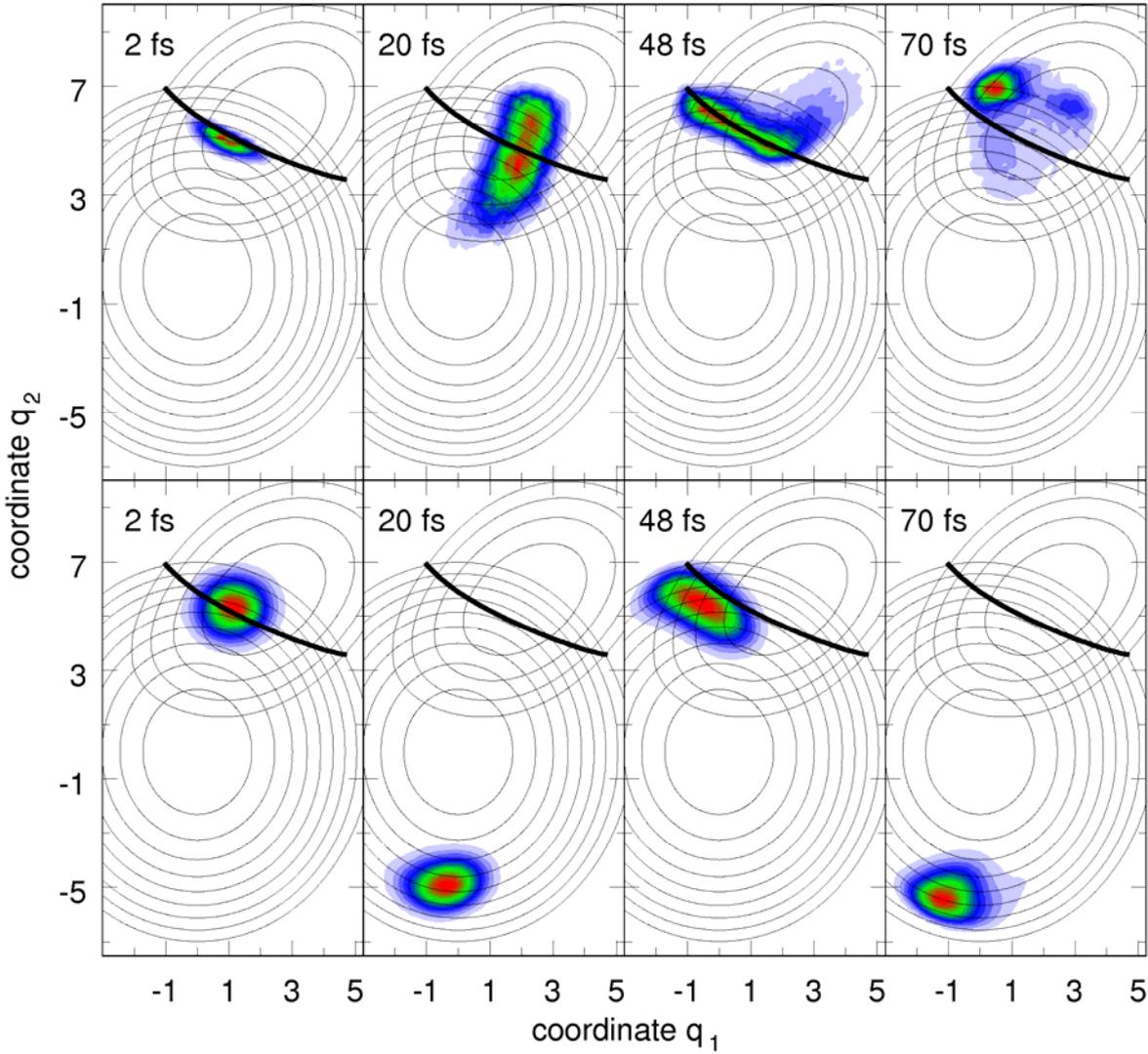